\documentclass[useAMS,usenatbib]{mn2e}

\usepackage{aas_macros,amsmath,graphicx,lineno,longtable,amssymb,mydefs}
\title[Transit Detection of a Starshade]{Transit Detection of a ``Starshade" at the Inner Lagrange Point of an Exoplanet}

\author[Gaidos et al.]{E. Gaidos,$^{1}$\thanks{Visiting Scientist, Center for Space and Habitability, University of Bern, Bern, Sitzerland. E-mail:  gaidos@hawaii.edu (EG)}\\
$^{1}$Department of Geology \& Geophysics, University of Hawaii at M\={a}noa, Honolulu, Hawaii 96822 USA\\
}

\begin{document}
\date{Submitted to MNRAS}
\pagerange{\pageref{firstpage}--\pageref{lastpage}} \pubyear{2016}

\maketitle

\label{firstpage}

\begin{abstract}
All water-covered rocky planets in the inner habitable zones of solar-type stars will inevitably experience a catastrophic runaway climate due to increasing stellar luminosity and limits to outgoing infrared radiation from wet greenhouse atmospheres.  Reflectors or scatterers placed near Earth's inner Lagrange point (\ellone{}) have been proposed as a ``geo-engineering" solution to anthropogenic climate change and an advanced version of this could modulate incident irradiation over many Gyr or ``rescue" a planet from the interior of the habitable zone.  The distance of the starshade from the planet that minimizes its mass is 1.6 times the Earth-\ellone{} distance.  Such a starshade would have to be similar in size to the planet and the mutual occultations during planetary transits could produce a characteristic maximum at mid-transit in the light-curve.  Because of a fortuitous ratio of densities, Earth-size planets around G dwarf stars present the best opportunity to detect such an artifact.   The signal would be persistent and is potentially detectable by a future space photometry mission to characterize transiting planets.  The signal could be distinguished from natural phenomenon, i.e. starspots or cometary dust clouds, by its shape, persistence, and transmission spectrum.
\end{abstract}


\section{Introduction}

Like every liquid water-covered planet around a solar-mass star, Earth has a serious greenhouse problem.  As the Sun converts hydrogen to helium and becomes denser, hotter, and more luminous \citep{Gough1981} the inexorable increase in incident irradiation on Earth will, all else being equal, cause surface temperatures to rise and the atmosphere to contain more water vapor.  Water is an efficient greenhouse gas, creating a positive feedback in Earth's climate system as more water vapor leads to elevated temperatures, and vice verse.  \citet{Walker1981} proposed that temperature-dependent aqueous weathering and precipitation of carbonate minerals acts as a negative climate feedback that adjusts atmospheric CO$_2$ to maintain weathering at a rate that balances volcanic degassing, i.e. surface temperatures permissive of abundant liquid water.     

However this planetary ``thermostat'' has its limits. As irradiance continues to increase, CO$_2$ will eventually disappears from the atmosphere causing a crisis for land plant life and any indeed any autotrophic life relying on atmospheric CO$_2$ as a source of carbon \citep{Caldeira1992}.  Beyond this point, increasing irradiance cannot be compensated by diminished CO$_2$ and temperatures rise.  Climate models predict an asymptotic (maximum) value for the outgoing infrared radiation of an Earth-like atmosphere as a function of temperature; if the absorbed incident radiation exceeds this value, temperatures will increase until the oceans evaporate \citep{Ingersoll1969}.  At that point multiple ``runaway" climate feed-backs, not mutually exclusive, can occur: water vapor in the upper atmosphere will be photolyzed and the escape of hydrogen will lead to loss of water.  Continued de-gassing of CO$_2$ from the mantle will not be compensated by aqueous weathering of silicates and formation of carbonates, enhancing the greenhouse effect and causing higher surface temperatures.   Eventually, surface temperatures will exceed the stability of carbonate minerals, leading to the breakdown of carbonate rocks in the crust, the release of $\approx$90 bars of CO$_2$ into the atmosphere \citep{Tuck1980}, a Venus-like climate, and the complete extinction of the biosphere.  Because of stellar luminosity evolution, all Earth-like planets within the ``conservative'' habitable zone of solar-mass stars \citep[0.95-1.67~AU][]{Kopparapu2013} will eventually orbit interior to the habitable zone and meet this fate.

A sufficiently capable intelligence (hereafter ``agency") could alter this course of events on an Earth-like planet, at least for a time.   Three possible options are to adjust the stellar output, alter the climate system of the planet, or modify the incident radiation on the planet.  The first is the most technically challenging, and the second could result in undesirable and/or unpredictable changes on the planet, while the third is more predictable and indeed possibly within reach of our civilization.  There are been several proposals to place reflecting or scattering material between the Earth and Sun, i.e. a ``Sunshade" at or near the first Lagrange point \ellone{} as a ``geoengineering'' solution to the less daunting (but more pressing) problem of anthropogenic climate change \citep[e.g.,][]{Govindasamy2000,Angel2006,Sanchez2015}.  The ``starshade" considered here would be designed to achieve an order of magnitude more attenuation with a lifespan of many Gyr.

Besides maintaining a temperate climate on a planet currently inside the habitable zone, an agency could ``resuscitate'' a planet orbiting interior to the inner edge of the habitable zone and which was undergoing a runaway moist greenhouse, but had not yet lost all its water nor had its carbonates destabilized.  Sufficient attenuation of incident radiation would allow re-precipitation of water, the resumption of silicate weathering, and stabilization of the climate under reduced irradiance.

The photometric transit technique is currently the most effective means of detecting Earth-size planets around solar-type stars.  A starshade at or near the \ellone{} point of a transiting planet will also periodically block a fraction of starlight reaching a distant observer.  Transits of a planet with a starshade could produce a unique lightcurve which could represent the signature of an extraterrestrial artifact, and this signature might be detectable with {\it current} means of transit detection.  Placement of the starshade near the \ellone{} point is an obvious choice, since this minimizes the requirements for station-keeping, but this point is dynamically unstable and elements must be either actively retained or replaced.  While this presents a challenge for any hypothetical engineer, it is a boon for an observer searching for signatures of intelligence, because any natural object would not persist at such a location for Gyr.  

The detection of intelligence via its manipulation of stellar emission is a venerable idea \citep[e.g., ``Dyson spheres"][]{Dyson1960}.  \citet{Arnold2005} showed how the transits of non-spherical artifacts could be identified by deviations in the lightcurves from those of planets.  For artifacts much smaller than the stellar disk, the  deviations are most pronounced during ingress and egress.   \citet{Korpela2015} examined the lightcurves produced by mirrors at the \elltwo{} point of a synchronously rotating transiting planet, i.e. around an M dwarf, placed there to illuminate the night side.  (See \citet{Lingam2017} for consideration of detection in reflected light).  The reflected light signature is too small to be detected by \kepler{} but might be detected by \jwst.  \citet{Wright2016} examined the information content of transit lightcurves as a means of identifying independently orbiting artifacts.      

\section{The Geometry of a Starshade}
\label{sec.geometry}

A starshade was assumed to consist of a ``cloud" of individual structures or particles that scatter or absorb incident starlight, but was modeled as a single continuous structure.  This structure was assumed to be a disk of of uniform effective opacity and radius $R_s$ centered on the planet-star axis.  More complex starshade geometries are considered in Sec. \ref{sec.discussion}. The attenuation of starlight averaged over the planet's surface and the transit light-curves of planet-starshade combinations were numerically calculated by decomposing both the planet and the starshade into finite elements with axisymmetric geometries.  This allows the effects of limb darkening on the star and scattering by the starshade to be explicitly included.  Limb darkening as a function of distance from the center of the stellar disk and effective temperature \teff{} was interpolated from the values tabulated by \citet{Claret2000}.    

\begin{figure}
\centering
   \includegraphics[width=\columnwidth]{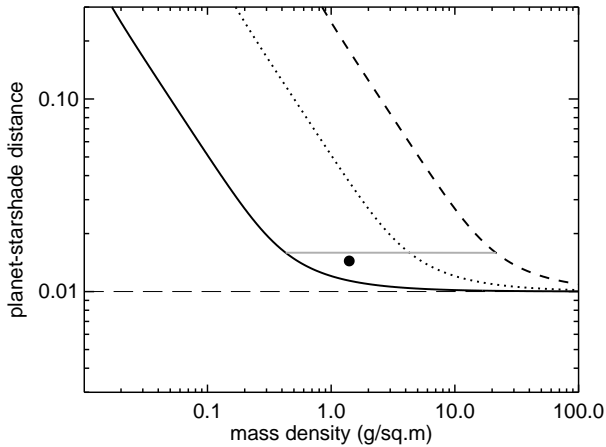}
   \caption{Distance between planet and starshade $x$ as a fraction of the planet-star distance for an Earth-Sun twin system, as a function of the mass surface density of the shade and for a momentum reflectivity $r$ of 1\% (solid), 10\% (dotted) and 50\% (dashed). The dashed horizontal line is the canonical \ellone{} point and the point is the nominal system described by \citet{Angel2006}.  The grey horizontal line is the locus of values that minimize starshade mass.}
\label{fig.distance}
\end{figure}

Due to non-gravitational forces, the elements of the starshade would -- in the co-rotating frame -- gyrate around a point at a fractional distance $x$ from the planet to the star near but not exactly at $x_1$, the position of \ellone{}.  \citet{Bewick2012} showed that the largest non-gravitational term by far is stellar radiation pressure.  The value of $x$ is given by the root of:
\begin{equation}
    \label{eqn.lagrange}
(1-x)^3 = 1 - \mu \frac{(1-x)^2}{x^2} - \frac{L_*r}{2\pi GM_* c \Sigma}(1-x)^2,
\end{equation}
where $\mu$ is the planet-star mass ratio, $L_*$ is the stellar luminosity, $G$ is the gravitational constant, $M_*$ is the stellar mass, $c$ is the speed of light, and $\Sigma$ is the mass per unit area of the star shade.  $r$ is the momentum reflectivity, i.e. the fraction of photon momentum impacted to the starshade relative to the purely reflecting case.  For an absorbing, non-reflecting surface, $r = 0.5$.  Fig. \ref{fig.distance} plots the distance of the starshade as a function of $\Sigma$ for $r = 0.01$, 0.1, and 0.5.  

For a fixed apparent angular size from the planet (i.e., for a given attenuation of irradiance), the physical diameter of a starshade scales as $x^2$ and its mass scales as $\Sigma x^2$.  For any value of $r$ there is a value of $\Sigma$ which minimizes the mass of the starshade \citep{Angel2006}.  This locus of minima is plotted as the grey line in Fig. \ref{fig.distance}.  This is $\approx1.6x_1$ and essentially independent of $r$ and $L_*$, and only weakly dependent on $M_*$ (through $\mu$).  It was assumed that the agency would adopt this mass-minimizing value and this was used in the remainder of this work.  The value of $\Sigma$ that minimizes total starshade mass, as well as the mass itself, depends on $r$.  

A lower $r$ allows the starshade to be placed closer to \ellone, and to be smaller and less massive.  \citet{Early1989} showed that the key to low $r$ is to scatter light at an angle just sufficient to miss the planet, rather than reflect it (this will be revisited in Sec. \ref{sec.phase}).  This can be accomplished by a perforated screen one half-wavelength thick that acts as a collection of half-wave plates with a narrow null along the forward-scattering (planet-ward) axis.  The nominal design studied by \citet{Angel2006} with $r = 0.026$ and $\Sigma = 1.4$~g~cm$^{-2}$ is plotted as the point in Fig. \ref{fig.distance}.  For 30\% attenuation and $r$ between 0.01 and 0.5, the minimum mass varies from  0.05 to 2.5 $\times 10^{12}$~kg, or about 1-50 times the mass of asteroid 25143 Itokawa visited by the {\it Hayabusa} spacecraft.

\begin{figure}
\centering
   \includegraphics[width=\columnwidth]{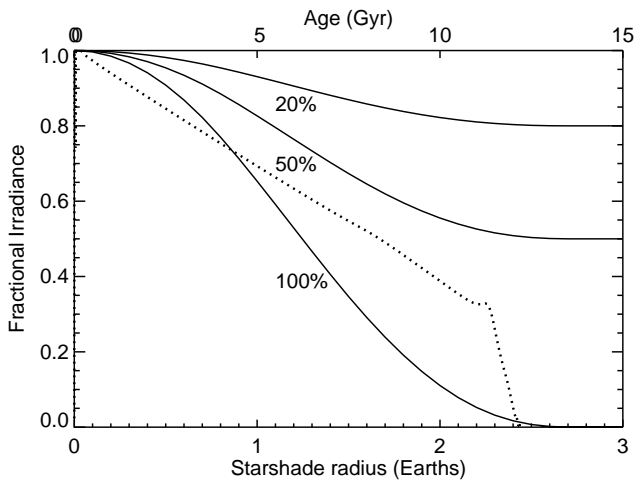}
   \caption{Irradiance experienced by an Earth-size planet around a Sun-like star, with a uniform, circular starshade at $1.6\times$ the distance to the \ellone{} point.  The irradiance is given as a fraction of the unattenuated value, for different starshade radii and opacities of 20, 50, and 100\%.  The dotted line is the irradiance reduction factor required to achieve the zero-age sequence value, as a fraction a function of age (top axis).}
\label{fig.shade}
\end{figure}

Larger and more opaque starshades block more of the incident stellar radiation -- to a point.  Figure \ref{fig.shade} plots the fractional irradiance of the planet in an Earth-Sun twin system with a starshade at the \ellone{} point vs. starshade radius for different values of starshade radius.  The star is entirely occulted if the starshade radius is $\gtrsim$2\rearth{}.  The dotted line in Fig. \ref{fig.shade} is the irradiance reduction factor required to achieve the zero-age main sequence (ZAMS) value at a given stellar age (top axis).  Thus to maintain ZAMS irradiance for 12~Gyr up to 70\% of incident starlight on the planet must be blocked or diverted. 

\section{Signatures of a Starshade}
\label{sec.signatures}

\subsection{Mutual Planet-Starshade Occultation}
\label{sec.occultation}

\begin{figure}
\centering
   \includegraphics[width=\columnwidth]{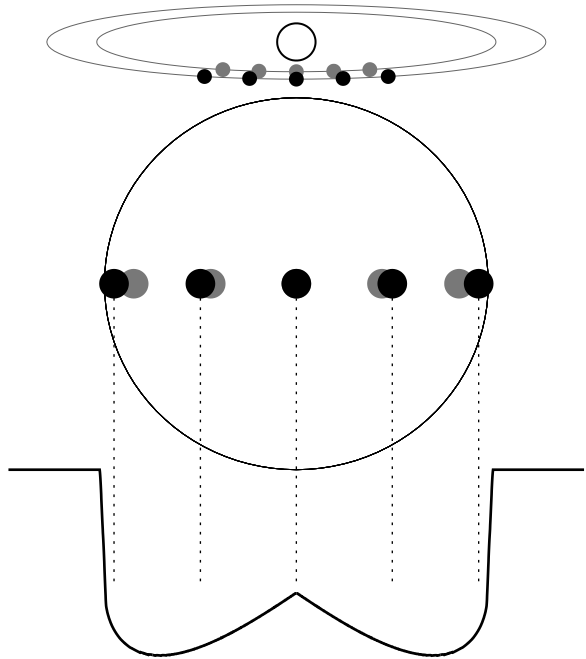}
   \caption{The geometry of a transiting planet (dark disk) with a starshade (grey disk) near \ellone.  The top figure is an oblique perspective, the middle figure illustrates the transit of the planet and starshade across the stellar disk, and the bottom figure is a normalized light-curve.  From the perspective of a distant observer, the planet lags the starshade at transit ingress, and the flux from the star is minimum.  The planet catches up and more fully occults the starshade at mid-transit, producing a local maximum, then leads the starshade, producing a second minimum.}
\label{fig.geometry}
\end{figure}

The geometry of a transiting planet and its starshade and the resulting light-curve are illustrated in Fig. \ref{fig.geometry}.    Due to its interior position at the \ellone{} point, ingress of the the starshade occurs first followed by the planet.  The planet's orbital velocity is greater than that of the starshade and it catches up at mid-transit and then leads the starshade at egress.  At ingress and egress, the projected separation of the two objects is maximum and thus the maximum occultation of the stellar disk (deepest transit) occurs.  At mid-transit the planet partially or completely occults the starshade producing a local maximum in the light-curve.  

\begin{figure}
\centering
   \includegraphics[width=\columnwidth]{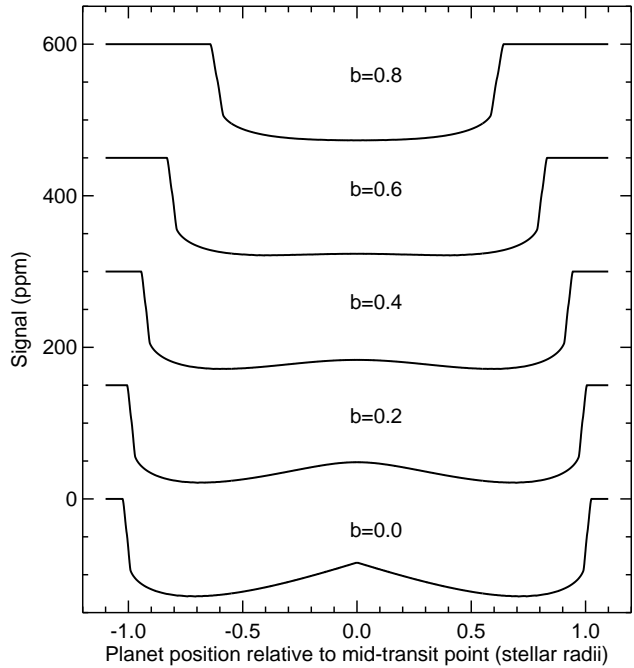}
   \caption{Predicted transit lightcurves of an Earth-size planet and equal-sized, completely opaque starshade, for different values of the transit impact parameter $b$.  The starshade is placed at $1.6\times$ the distance of the \ellone{} point from the planet.  The response function is that of \kep{} and the host star is solar-type.}
\label{fig.lightcurve}
\end{figure}

Greatest planet occultation of the starshade and the most pronounced mid-transit maximum occurs when the transit impact parameter $b$ (the distance between mid-transit point and center of the stellar disk) is zero.  For $b>0$, the observer is above/below the orbital plane, the starshade is less occulted by the planet, and the maximum is smaller or non-existent.  Figure \ref{fig.lightcurve} shows lightcurves for an Earth-Sun-twin system with a completely opaque starshade with radius 1\rearth{} and different values of $b$.    

\begin{figure}
\centering
   \includegraphics[width=\columnwidth]{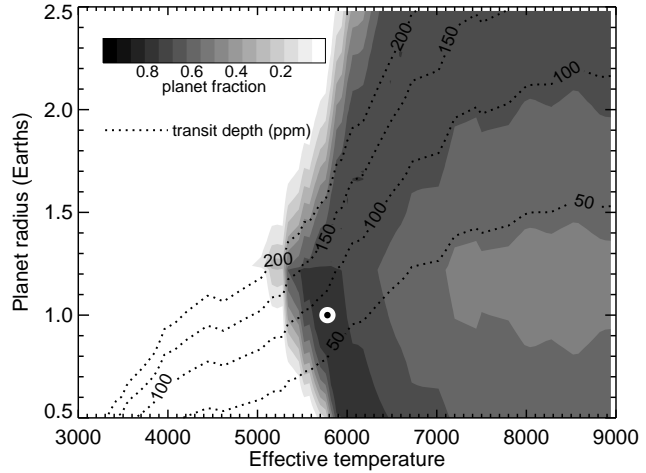}
   \caption{Fraction of planets meeting the criteria for planet-starshade mutual occultation (Eqns. \ref{eqn.criterion1} and \ref{eqn.criterion2}).  Contour lines of equal transit depth are also plotted and the Sun-Earth system is marked.}
\label{fig.criterion}
\end{figure}

The \ellone{} point is at a fractional distance $(\mu/3)^{1/3}$ from the planet towards the star.  From the perspective of a distant observer, this displacement has a maximum value at transit ingress/egress $\approx R_*(\mu/3)^{1/3}$.  The starshade will be mostly unocculted by the planet at this time if $R_*(\mu/3)^{1/3} \gtrsim R_p$.  This condition can be expressed in terms of mean densities:
\begin{equation}
\label{eqn.criterion1}
\rho_p \gtrsim 3\rho_*.
\end{equation}
Since $\rho_{\oplus} = 3.9\rho_{\odot}$ this criterion is satisfied for an Earth-Sun-like system.  If the transit occurs with impact parameter $b$ then significant occultation of the planet-shade will occur at mid-transit if
\begin{equation}
\label{eqn.criterion2}
b < \left(\frac{3\rho_*}{\rho_p}\right)^{1/3}.
\end{equation}
The fraction of planets that satisfy this criterion as a function of planet radius and stellar effective temperature is plotted in Fig. \ref{fig.criterion}.  Planet masses were calculated from radii by sampling the statistical relations of \citet{Chen2017}.  Stellar masses and radii were interpolated from values tabulated in \citet{Pecault2013}.  Impact parameter was uniformly distributed over $0 < b < 1$.  

The regime in which most starshades could be most readily detected occupies a small fraction of this star-planet parameter space, but it fortuitously includes Earth-size planets around G dwarfs.  M and K dwarfs are not suitable targets for starshade searches because they are too small compared to the size of the planets\footnote{The slow evolution of these stars also eliminates one rationale for constructing such a starshade.}.  Starshades around hotter stars satisfy Eqns. \ref{eqn.criterion1} and \ref{eqn.criterion2} but it is more difficult to detect Earth-size planets and starshades of similar size around them.  The break at 1.2\rearth{} is a product of the piece-wise mass-radius relation of \citep{Chen2017}

\begin{figure}
\centering
   \includegraphics[width=\columnwidth]{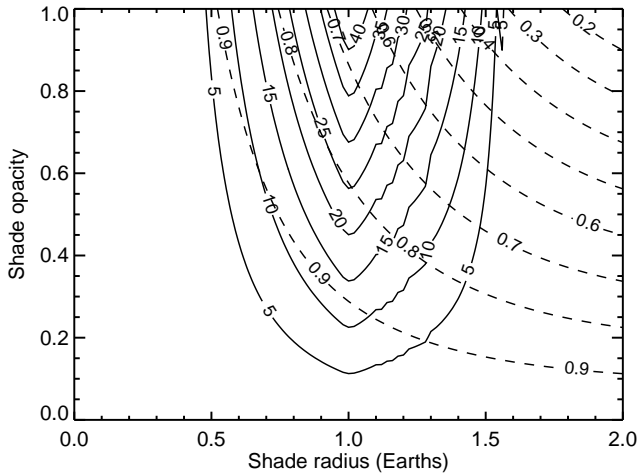}
   \caption{Contours plot of the difference in ppm between the mid-transit maximum and the absolute minimum in the lightcurves of transiting planets with starshades (solid lines).  The planet and starshade are both Earth-size, the starshade is opaque and is placed at $1.6\times$ the distance of the \ellone{} point from the planet, the host star is solar-type, and $b = 0$.}  
\label{fig.solar}
\end{figure}

A distinguishing signature of the starshade at \ellone{} is a local maximum in the light-curve at mid-transit, where mutual planet-starshade occultation is greatest, with two leading and lagging minima (Fig. \ref{fig.lightcurve}).  Figure \ref{fig.solar} plots contours of constant difference (ppm) in the normalized flux between the minima on either side of the mid-transit point and the mid-transit maximum vs. starshade radius and opacity (solid lines) for $b=0$.  Also plotted are contours of constant planet irradiance as a fraction of the unimpeded value (dashed lines) vs. shade radius and opacity.  A starshade smaller than a minimum size does not produce a local maximum because the effect of mutual occultation is offset by limb darkening.  A much larger starshade also does not produce the characteristic double-minima because the planet occults some part of the starshade during the entire transit event.  For $b=0$, a starshade equal in size to the planet produces the maximum effect.  This shifts to larger starshades for $b>0$.   For a given attenuation of stellar flux, the signal produced by a starshade will depend on its size or opacity.  

\subsection{Ingress/Egress Duration}
\label{sec.ingress}

\begin{figure}
\centering
   \includegraphics[width=\columnwidth]{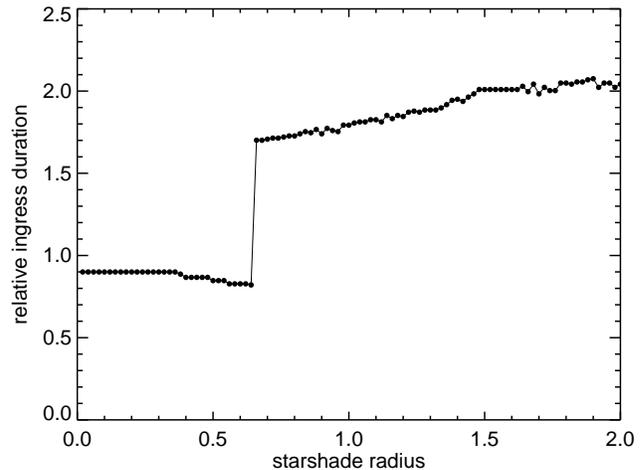}
   \caption{Duration of ingress/egress of a transiting exoplanet plus starshade compared to the theoretical value for the planet only.  The planet and starshade are both Earth-size and orbit a solar-type star; the starshade is completely opaque and is at $1.6\times$ the planet-\ellone{} distance.}
\label{fig.ingress}
\end{figure}

The projected offset of the starshade with respect to the planet from the perspective of a distant observe will prolong ingress and egress compared to the planet-only case.  Figure \ref{fig.ingress} shows ingress/egress duration $\tau$ as a fraction of the transit duration, for the nominal Earth-Sun-twin case.  First and last contact during ingress/egress correspond closely to inflection points in the light-curve, and ingress/egress duration was calculated as the interval between the local maximum and minimum in the second derivative.  $\tau$ calculated in this manner is  not affected by small starshades $R_s < 0.6$\rearth{} but nearly doubled for larger ones.    

For the planet alone, the ratio of the ingress/egress duration $\tau$ to transit duration $T$ \citep[e.g.,][]{Seager2003} is
\begin{equation}
\frac{\tau}{T} = \frac{1}{1-b^2}\frac{R_p}{R_*}.
\end{equation}
$R_p/R_*$ can be independently determined from the transit depth $\delta$.  $b$ can be determined by its relation to the transit duration $T$, $\rho_*$, the orbital eccentricity $e$ and longitude of periastron $\omega$:
\begin{equation}
\label{eqn.duration}
1-b^2 = \sqrt{\delta}\left(\frac{3P}{\pi G \rho_* T^3}\right)^{2/3}\frac{1-e^2}{\left(1+e\cos\omega\right)^2}.
\end{equation} 
If $\rho_*$ can be independently determined, e.g. by asteroseismic analysis of oscillations, and the planet is known to have a near-circular orbit, e.g. by Doppler radial velocity observations, then a comparison between the observed transit duration and that given by Eqn. \ref{eqn.duration} could reveal a starshade.  Normally, $b$ could be estimated directly from fitting the shape of the transit light-curve with a model that includes limb darkening constrained by known properties (i.e., \teff{} and $\log g$) of the host star. However, the mutual occultation of the starshade will cause the light-curve to be flatter and thus $b$ could be underestimated and thus $\tau$ overestimated.   The starshade also increases $\delta$ and thus $\tau$.

\subsection{Occultation by a Second Planet}
\label{sec.occultations}

Approximately 40\% of \kep{} candidate planets are in multi-planet systems\footnote{https:$\backslash\backslash$exoplanetarchive.ipac.caltech.edu} and statistical analyses of these systems show the mutual inclinations to be small $\lesssim1^{\circ}$ \citep{Lissauer2011,Figueira2012,Fang2012,Gaidos2016}.  Mutual occultation will occur if the planets transit each other on the disk, producing a more complex lightcurve with multiple minima \citep{Sato2009,Pal2012}.  The presence of a starshade could be revealed if the second planet was occulted by the starshade as both transit the stellar disk.

\begin{figure}
\centering
   \includegraphics[width=\columnwidth]{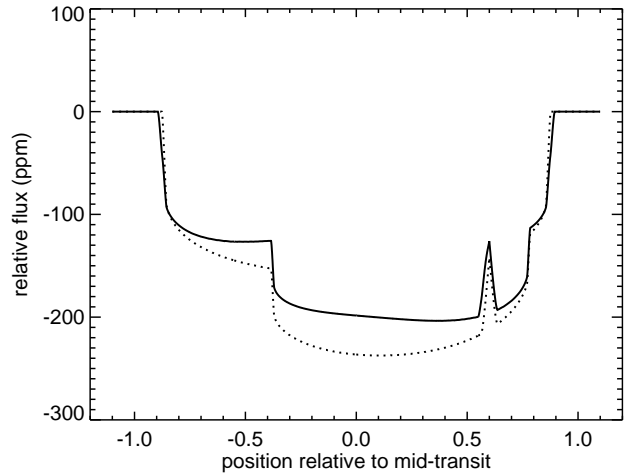}
   \caption{Light curve produced by two transiting Earth-size planets, with the starshade of the exterior planet occulting the interior planet (solid curve).  The dashed curve is the same scenario except the starshade is removed and the outer planet radius is increased by $\sqrt{2}$.  In each case the lightcurve contains a peak produced when the inner planet is occulted by the outer planet, plus starshade, if present.}
\label{fig.twoplanets}
\end{figure}

Figure \ref{fig.twoplanets} shows the lightcurve for a scenario where a second, Earth-size inner planet overtakes the planet with an opaque starshade of the same size and is occulted by both (solid line).  The individual transits are seen, plus the peak where the inner planet is occulted by outer planet, plus its starshade.  The dashed line is the lightcurve for the same scenario, minus the starshade, but with the outer planet increased in radius by $\sqrt{2}$.  The lightcurve is similar in shape and magnitude, and the peak remains since the outer planet still occults the inner planet.  The similarity of the lightcurves means that it would be difficult to distinguish between these scenarios and thus uniquely identify a starshade.  The peak could be a signature if it was known a priori that the interior planet was not occulted by the outer planet, but the transit geometry will not be this precisely known.     

Moreover, the probability of such fortuitous geometries is low.  Consider two planets on circular, near edge-on orbits with a small mutual inclination. The projected orbits of inner planet (1) and outer planet (2) will intersect if
\begin{equation}
\lvert a_1\left(\cos\Omega+\phi\sin\Omega\right)(i_1-i_2)-\left(a_2-a_1\right)i_2\rvert < R_1 + R_2
\end{equation}
where $a$ is the semi-major axis, $\Omega$ is the ascending node of the inner orbit relative to the plane of the outer orbit, and $\phi$ is the orbital phase relative to inferior conjunction.  If $R_* \ll a$ then $\phi \ll 1$ and, as long as $\cos \Omega$ is not small, the second term on the left hand side can be neglected.  The probability of an occultation is equal to the product of the probability density functions (PDFs) of $i_2$ and $\delta \equiv i_1-i_2$ integrated over a narrow region along the line 
\begin{equation}
\label{eqn.delta}
\delta = \frac{(r-1)i_2}{\cos \Omega},
\end{equation}
where $r = a_2/a_1 > 1$.  The two-dimensional integral in this narrow region is converted to a one-dimensional integral over $i_2$ multiplied by the height $2(R_1+R_2)/a_1\cos \Omega$.  Taking the PDFs for $i_2$ and $\delta$ to be uniform over $0 < i_2 < a_2/R_*$ and a Rayleigh distribution with dispersion $\sigma$, respectively, the probability becomes   
\begin{equation}
\label{eqn.prob1}
p = \frac{1}{2\pi}\int_0^{2\pi} d\Omega \int_0^{R*/a_2} di_2\frac{a_2}{R_*}\frac{\delta}{\sigma^2}e^{-\delta^2/2\sigma^2}\frac{2\left(R_1+R_2\right)}{a_1 \cos \Omega}.
\end{equation}
Using Eqn. \ref{eqn.delta}, Eqn. \ref{eqn.prob1} becomes
\begin{equation}
\label{eqn.prob2}
p = \frac{r(R_1+R_2)}{\pi R_*(r-1)} \int_0^{2\pi} d\Omega \int_0^{\frac{R*(r-1)}{a_2 \cos \Omega}} d\delta \frac{\delta}{\sigma^2} e^{-\delta^2/2\sigma^2}.
\end{equation}
The inner integral is readily evaluated:
\begin{equation}
\label{eqn.prob3}
p = \frac{r(R_1+R_2)}{\pi R_*(r-1)} \int_0^{2\pi} d\Omega \left[1 - {\rm exp}\left(-\frac{R_*^2(r-1)^2}{2\sigma^2a_2^2\cos^2\Omega}\right)\right].
\end{equation}
For $R_* \ll a_2$ most of the contribution to the integral comes where $\cos \Omega \ll 1$.  Exploiting the symmetry of the integrand around $\Omega = 0$ and using an appropriate substitution of variables;
\begin{equation}
\label{eqn.prob4}
p \approx \frac{2\sqrt{2}r}{\pi \sigma}\frac{R_1+R_2}{a_2}\int_0^{\infty}dz\,z^{-2}\left(1-e^{-z^2}\right).
\end{equation}
The definite integral has the value $\approx 1.77$.  Expressing $\sigma$ in degrees, $R_1$ and $R_2$ in Earth radii and $a_2$ in AU, then
\begin{equation}
p \approx 0.004 \frac{r}{\sigma}\frac{R_1+R_2}{a_2}
\end{equation}
For a pair of Earth-size planets with $a_2 = 1$, $r = 1.5$, and $\sigma = 1^{\circ}$ \citep{Fang2012} the probability of mutual occultation is about 1\%.  However, the integrals in Eqn. \ref{eqn.prob1}-\ref{eqn.prob4} were not conditionally evaluated with the inner planet actually transiting, so the fraction of  multi-transit systems that undergo mutual occultations will be somewhat higher.  The probability will also be higher if $a_2 < 1$~AU, as is the case for nearly all \kep{} systems.

\subsection{Anomalous Mass-Radius}
\label{sec.massradius}

A starshade can block a significant fraction of stellar flux compared to its negligible mass, thus the combination of planet plus shade could appear as a planet with an anomalously large radius for its mass.  However, significant scatter appears to be intrinsic to the mass-radius relationship for Earth-size planets \citep{Wolfgang2016}.  Assuming that the mass of an Earth-size planet could be measured with an error of 25\%, the statistical relation of \citet{Chen2017} predicts a dispersion in $R_p$ of 8.5\%, neglecting errors in stellar radius.  For comparison, the presence of a starshade of equal size increases the transit depth by up to 45\%, depending on the impact parameter, and the apparent radius by up to 20\%.  Thus a planet with a starshade may appear as a low-density outlier among Earth-size planets. 

\subsection{Phase Angle and Spectral Effects}
\label{sec.phase}

\begin{figure}
\centering
   \includegraphics[width=\columnwidth]{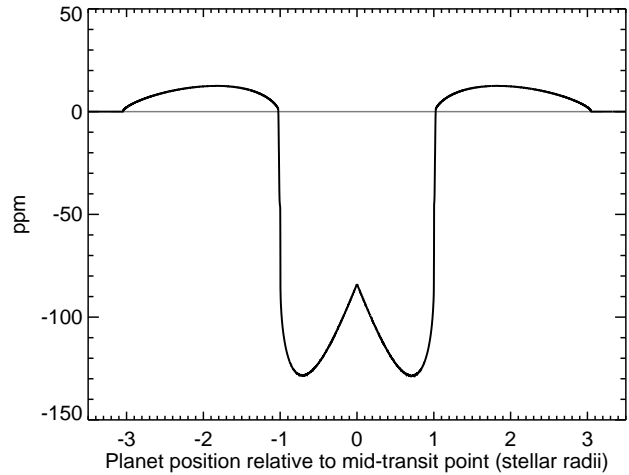}
   \caption{Light-curve of a 1\rearth{} planet with an equal-sized starshade that scatters all light at angle $2R_*/a$.  The host star is sun-like and the transit impact parameter $b = 0$.}
\label{fig.scatter}
\end{figure}

A well-engineered starshade will scatter light at an angle
\begin{equation}
\label{eqn.angle}    
\theta > \frac{R_p}{xa} + \frac{R_*}{a} \approx 1.6\frac{R_*}{a}
\end{equation}
that avoids the planet but minimizes the transfer of momentum to the starshade (proportional to $\theta^2$) and thus minimizes the required mass \citep[][Fig. \ref{fig.distance}]{Early1989}.  This will contribute to the lightcurve at phases outside of transit.  The irradiance at the starshade scales as $(R_*/a)^2$, which is only $10^{-4}$ but the solid angle into which the light is scattered scales as $\theta^2 \sim (R_*/a)^2$ (Eqn. \ref{eqn.angle}), thus these factors cancel.  The maximum potential signal can be estimated by assuming it is all symmetrically scattered at a single angle.  Figure \ref{fig.scatter} shows the result for $\theta = 2R_*/a$ or about 0.5~deg for an Earth-Sun-like system.  Scattering produces 10~ppm ``shoulders" before and after transit.  

An alternative to macroscopic elements for a starshade is dust.  A silicate dust cloud produced from an asteroid has been proposed for geoengineering of terrestrial climate because of its high surface area-to-mass ratio and comparative simplicity of production \citep{Struck2007,Bewick2012}.  However, a massive source body will orbit at \ellone{}, exterior to the equilibrium point of the dust (Fig. \ref{fig.distance}).  As a consequence, dust will be rapidly dispersed and must be continuously replaced.   Quasi-spherical dust grains are also less efficient scatterers per unit mass than thin interference films and can strongly forward-scatter, reducing their ability to shade the planet.  Regardless, the phase function for Mie scattering is set by the particle size parameter and the solid angle of scattering will not scale as $(R_*/a)^2$.  This means the scattered light from a dust cloud will be negligible compared to line-of-sight extinction it produces, and the transit light-curve will appear similar to that of one composed of macroscopic elements, depending on the projected dust profile.   

\begin{figure}
\centering
   \includegraphics[width=\columnwidth]{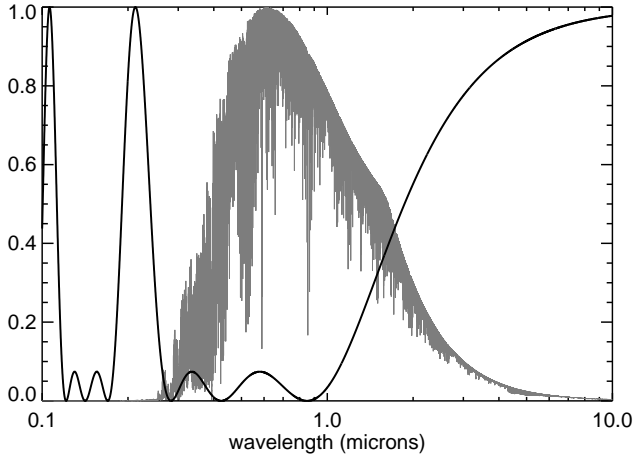}
   \caption{Transmission function of the thin-film sunshade proposed by \citet{Angel2006}.  The spectrum of a solar-type star based on the PHOENIX code is plotted for reference.}
\label{fig.transmit}
\end{figure}

An engineered starshade will probably be achromatic, i.e. optimized only to deflect light at wavelengths were there is significant stellar flux.  The on-axis transmission function of the interference screen design proposed by \citet{Angel2006} was calculated using his Eqn. 1 and plotted in Fig. \ref{fig.transmit}.  The spectrum of a solar-type star calculated using the PHOENIX stellar atmosphere code \citep{Hauschildt1997} is also shown.  Light at wavelengths around the peak of solar emission are efficiently scattered, while those at $\lambda > 2\mu$m and near $0.2\mu$m are not.  The exact response function depends on the detailed structure of the film, but transparency at wavelengths $\lambda \gg 1\mu$m will be an inevitable consequence of the need to minimize mass and the relatively minor contribution of this wavelength range to the total stellar irradiance from a solar-type star.   Thus such a starshade will appear significantly larger/smaller at shorter/larger wavelengths.

\begin{figure}
\centering
   \includegraphics[width=\columnwidth]{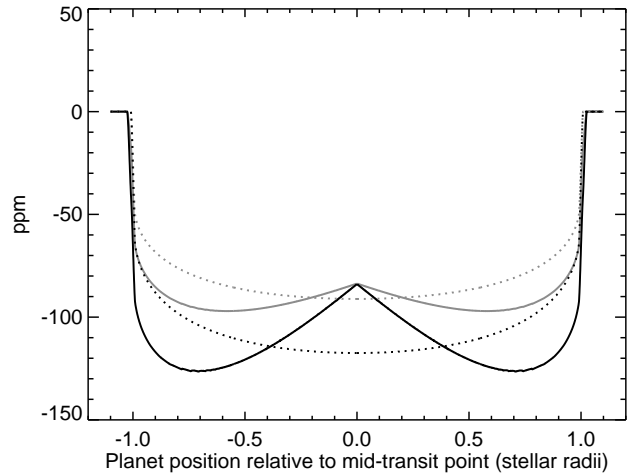}
   \caption{Predicted transit lightcurves in the \kep{} band-pass (black lines) and infrared $K$-band (grey lines).  Solid lines are for a planet with a thin atmosphere (neglected) and thin-film starshade as described by \citet{Angel2006}.  Dotted lines are for a planet with a solar-metallicity atmosphere at 255~K containing a layer of 0.05 $\mu$m haze particles at 1 $\mu$bar.}
\label{fig.colorlc}
\end{figure}

The transit depth of a planet with a starshade having the thin-film design described by \citet{Angel2006} was calculated for both \kep{} ($K_p$) and 2.2$\mu$m $K$-band-like response functions.  The planet and starshade have radii of 1\rearth{} and the host star has a solar-type spectrum.  The transmission averaged over the $K_p$ and $K$ band-passes are 4\% and 60\%, respectively.  The two light-curves are plotted in Fig. \ref{fig.colorlc}, showing that the starshade is nearly invisible in $K$-band.

\subsection{Infrared Emission}
\label{sec.infrared}

Thermal emission from an unresolved transiting planet (and any starshade) can be detected by observing it if and when it is eclipsed by its host star.  An electromagnetically inefficient starshade that absorbs, rather than scatters light, will reach a temperature $T_s$ that is close to or in excess of that of the planet $T_p$;
\begin{equation}
    T_s = T_p\left(\frac{2}{\epsilon(1-x)^2}\frac{1-A_s}{1-A_p}\right)^{1/4},
\end{equation}
where $A$ is the albedo and $\epsilon$ is the starshade emissivity (that of the planet was assumed to be unity).  The maximum emission will occur from a black-body starshade ($\epsilon \approx 1$, $A_s \approx 0$); in front of an Earth twin ($T_p = 255$K, $A_p = 0.31$) this object would radiate at $T_s = 335$K.

Besides low albedo, a major reason for the hotter starshade is the factor of $2^{1/4} \approx 1.2$ that arises from its disk geometry vs. the spherical geometry of the planet and efficient heat redistribution by oceans and atmosphere.  If the radii of the planet and starshade are similar, the combined spectra closely resemble a slightly-larger black-body-like planet at 325K.  The maximum equilibrium temperature of a planet with an Earth-like orbit/irradiation and efficient heat distribution but arbitrarily low albedo is 280K.  Thus if it can be independently established that the planet has a thick atmosphere and efficient heat transport, but appears to radiate well above 280K, then extra emitting area in the planet's vicinity could be suspected.   However, black-body absorbers are a very inefficient design for a starshade (Sec. \ref{sec.geometry}), and if the irradiance is attenuated by scattering (Sec. \ref{sec.phase}), rather than absorbing the starshade will be much colder and make negligible and undetectable contribution to the emission at infrared wavelengths.   

\section{Detection}
\label{sec.detection}

To detect the mid-transit maximum in a light-curve due to a starshade with the equivalent of $5\sigma$ significance, the photometric signal-to-noise ratio (SNR) must be $>5.5$ ($\nu = 2$ degrees of freedom) over the ingress to mid-transit interval (6.5~hr for planet on an Earth-like orbit transiting a solar-type star observed at low $b$).  Therefore to detect the maximum signal of 44 ppm in Fig. \ref{fig.solar}, the photometric precision must be $< 8$~ppm.  For comparison, the best 10th percentile precision for 6~hr achieved by \kep{} for observations of very bright stars is 12 ppm \citep{Christiansen2012}, while the design performance of the CHaracterizing ExOPlanets Satellite \citep[CHEOPS,][]{Broeg2014} will be 20 ppm.  These are for single transits, and observations of multiple transits by a \kep{}-like observatory could put the necessary precision within reach.  

On the other hand, noise due to granulation and magnetic activity on solar-type stars imposes a 6~hr noise ``floor" of about 10 ppm \citep{Basri2013}.  Better precision might be obtained in the infrared where such variability is expected to be less, and the \jwst{} will have far greater sensitivity \citep{Beichman2014}.  Slightly larger starshades around slightly cooler (late G-type) stars would provide up to twice the signal, but the condition for mutual occultation ultimately limits the search space in the Hertzsprung-Russell diagram (Fig. \ref{fig.criterion}).

Detecting the prolongation of ingress or egress duration by a starshade might be feasible with high cadence, high SNR photometry.  The duration for the Earth-Sun case is at least 7 min, depending on $b$.   Given sufficient SNR, it could be resolved in observations with the 3~min short cadence of \kep{} and the 2~min short cadence of \tess{} \citep{Ricker2014}.  The error in the predicted $\tau$ (Eqn. \ref{eqn.duration}) will come from the determination in density (a few percent) \citep{Huber2013} and from the measurement of eccentricity, which, even if Doppler measurements are available, will still be $\pm 0.1$ even if eccentricity is low.  While large star-shades ($\gtrsim 2R_p$) may be apparent in ingress/egress photometry, these ambiguities remain challenges for this detection strategy.

The best prospects for the detection of a starshade might be transit spectroscopy or spectrophotometry, at least for the class of low-mass, thin-film shades that scatter at visible wavelengths but are mostly transparent in the near-infrared.  The prospective observational portfolios of the next generation of Extremely Large Telescopes (ELTs) such as the Thirty Meter Telescope, the European-ELT, and the Giant Magellan Telescope include transit spectroscopic searches for atmospheres of Earth-size planets \citep{Snellen2015,Crossfield2015,Lovis2017}. 

\section{Potential False Positives}
\label{sec.fp}

The transit of a planet across a starspot will also produce a light-curve with a local maximum \citep[e.g.,][]{Pont2007,SilvaValio2008,Nutzman2011} that, if near mid-transit, could be mistaken for a starshade occultation.  However spots are transient features on the stellar surface and all stars rotate; starspot-produced features will change or disappear during successive transits, while a starshade will persist.  Likewise, mutual occultations of natural satellites could also produce features in the transit lightcurve \citep{Kipping2012} but these events will not occur consistently at the transit mid-point and satellites and their signals are likely to be much smaller than the planet (0.27 for the Moon).   

A natural dust cloud produced by the collision or rotational breakup of an asteroid or comet nucleus could also exist on an interior orbit, but the lifetime of micron-size dust particles to Poynting-Robertson drag within 1~AU is $10^3-10^4$~yr and dust cannot be permanently captured in any of the equilibrium points \citep{Liou1995}, and thus the dust must be continuously replenished.  A cometary body could in principle be the dust source, but its orbit would not be stable on long time scales either.  Instead, dust would disperse into clumpy rings and form ``exozodiacal" clouds \citep{Stark2013}.

Mie scattering by a high-altitude haze in the atmosphere of a transiting planet increases the apparent radius at shorter wavelengths, mimicking the wavelength-dependent opacity of a thin-film starshade.  For example, the transmission function of the \citet{Angel2006} shade (Fig. \ref{fig.transmit}), which rolls over at $\lambda > \approx 0.85\mu$m, can be crudely reproduced by a haze of $0.85/(2\pi) \approx 0.14\mu$m particles.  The magnitude of the difference between the transit depth in the optical ($\lambda < 0.85\mu$m) and infrared ($\lambda > 0.85\mu$m) is also similar to that produced by a plausible atmosphere.  Transit depths in the \kep{} and infrared $K$ band-passes were estimated for a hypothetical Earth-size planet with a solar-composition atmosphere (mean molecular weight of 2.3) an equilibrium temperature of 255~K, and a haze layer at $1\mu$bar composed of 0.1$\mu$m particles.  A  simulation of Kepler-289d with $T_{\rm eq}=1000$K \citep{Gaidos2017} was used and the spectral features scaled to Earth by the factor $H/R_p$, where $H$ is the atmosphere's scale height, a factor of 0.4 in this case.  

The predicted transit lightcurves are plotted as the grey solid and dotted lines in Fig. \ref{fig.colorlc}; the difference is comparable to that of a starshade.  On the other hand, a hazy atmosphere does {\it not} produce a mid-transit peak in the light curve (Fig. \ref{fig.colorlc}), and the unobscured infrared spectrum of such an atmosphere will contain the absorption features of any major molecular constituents, e.g. H$_2$O and CO.  Larger haze particles could blanket such features, but these would also remove the difference between the apparent radius at short and long wavelengths.      

Although an anomalously high emitting temperature from an Earth-like planet could be due to the presence of additional emitting area, e.g. a starshade, it could alternatively be explained by inefficient heat redistribution around the planet.  An Earth-like planet could also possess a Moon-like natural satellite.  Due to the lack of an atmosphere and its low albedo, daytime temperatures on the Moon can exceed 387K \citep{Williams2017} and its infrared flux makes a significant contribution to variable signal from the Earth-Moon system \citep{Moskovitz2009}.  

\section{Discussion}
\label{sec.discussion}

An artifact placed near the \ellone point of an Earth-like planet to modulate stellar irradiance and maintain a habitable climate against stellar evolution or move a planet into the habitable zone could be detected by transit photometry or spectroscopy.  Mutual occultation of the planet and starshade during a transit will produce a characteristic local maximum in the light-curve at the mid-transit point, as long as the impact parameter is small.  These occultations will be most detectable for rocky Earth- and super-Earth-size planets ($R_p < 1.2$\rearth{}) around solar-type stars.  The required photometric stability to detect such features is probably beyond the reach of current space-based observatories, but might be reached by \jwst{} or potential future missions such as LUVOIR \citep{Crooke2016}.  The use of other purely photometric signatures of a starshade suffer from imprecisely known orbital elements (transit duration), low probability (occultations by a second planet) or an incomplete picture of planet properties (anomalously high radius for a measured mass).  A less ambiguous signature is a decreasing transit depth wavelength combined with the lack of molecular features expected if this was due to a low molecular-weight atmosphere.  Transmission spectroscopy or spectrophotometry with instruments on ground-based ELTs might detect such objects as part of a survey for atmospheres of Earth-like planets. 

A starshade might be indirectly detected via its effect on the planet's energy balance, e.g. an equilibrium temperature lower than expected for known stellar irradiance and planetary albedo.  Independently establishing a planetary albedo would require the very difficult task of detection at visible wavelengths.  Signatures of liquid water oceans \citep[e.g.,][]{Williams2008} from a planet orbiting at a distance where the stellar irradiance is above the greenhouse limit might also suggest flux modification, but again the planetary albedo must be known.  Moreover, an unresolved starshade could make the planet appear hotter than it actually is (Sec. \ref{sec.infrared}).      

As the host star evolves along and then off the main sequence, it will become larger and hotter until reaching the turn-off where it expands and \teff{} drops, but luminosity continues to increase along the red giant branch.  Barring significant mass loss and orbital evolution of the planet, this will cause two effects: the star will subtend a larger angular diameter as seen from the planet, and increasing radiation pressure from the star will move the equilibrium point of a starshade of a given mass density further from the planet.  A starshade of fixed physical size maintained at the equilibrium point will become less effective, at the same time it must become {\it more } effective to maintain a constant irradiance of the planet.         

\begin{figure}
\centering
   \includegraphics[width=\columnwidth]{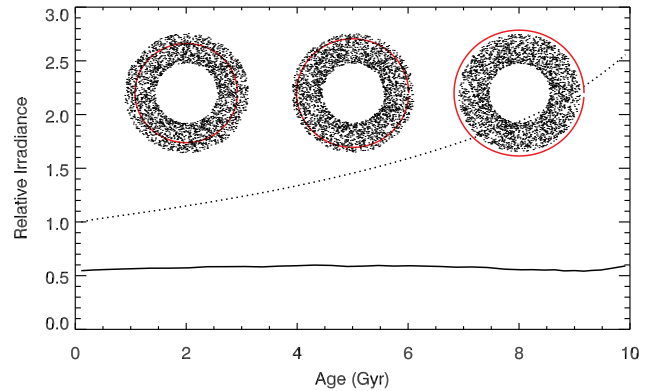}
   \caption{Geometry of a starshade that passively maintains constant irradiance of a planet against 10~Gyr of evolution of its solar-mass host star.  The illustrations are three ``snapshots" of the starshade and the star as seen from the planet.  The dotted and solid curves are the unmitigated and modulated irradiance values, respectively.}
\label{fig.evo}
\end{figure}

The starshade could possess some active mechanism to reconfigure itself over billions of years as the star evolves, or its geometry could passively maintain a near-constant irradiance by design.  Perhaps the simplest geometry is an annular one where only the central disk of the star is un-obscured, and expansion of the star has a reduced effect on irradiance.  Fig. \ref{fig.evo} illustrates such a geometry and the unmitigated and modulated irradiance values are plotted vs. stellar age as the dashed and solid lines.  A solar mass, solar metallicity evolutionary track from the Dartmouth Stellar Evolution Program \citep{Dotter2008} was used to calculate the irradiance.  The transit lightcurves for the geometry shown in Fig. \ref{fig.evo} is qualitatively similar to the nominal case (Fig. \ref{fig.lightcurve})  because the starshade, like the planet, is much smaller than the stellar disk from the perspective of a distant observer.  However, the starshade and planet are of similar size, and if the starshade aperture is large enough the mid-transit peak in the lightcurve can disappear or even invert.  

The caretakers of a planet orbiting a rapidly-evolving high-mass star have more cause to construct a starshade than those orbiting a low-mass star, especially an M dwarf, although ``resuscitation" of a planet would be an exception.  This kind of reasoning also favor searches for starshades around {\it older} stars, assuming that the initial (zero-age) stellar irradiance for a planet must be below the greenhouse limit.  

Instead of modifying the incident flux at the top of the atmosphere, the agency could elect to increase the planet's albedo by adding scattering particles to the atmosphere\footnote{\citep{Gerstell2001} proposed climate engineering in the opposite sense, i.e. the addition of ``super-greenhouse" gases that could also be remotely detected.}.  Addition of sulphate aerosols to Earth's stratosphere has been considered as a means to offset radiative forcing by anthropogenic CO$_2$ \citep{Govindasamy2000,Rasch2008,Pierce2010}, but this can have undesirable climatic and chemical effects \citep[e.g.,][]{Heckendorn2009,Ferraro2011}.  The ideal particles would be efficient scatterers at the peak wavelength of the stellar radiation, but inefficient scatterers at the longer wavelengths where the planet emits, and they would be long lived, and chemically and biologically neutral \citep{Dykema2016}.  In principle, hazes alone can drive the planet's spherical albedo towards unity \citep{VandeHulst1974} providing sufficient leverage to compensate the luminosity evolution of a Sun-like star on the main sequence.

The presence of hazes or cloud droplets in the atmospheres of exoplanets has been inferred by the suppression of gas-phase absorption features and/or a negative slope due to Rayleigh scattering regime at short wavelengths in spectra obtained during transits \citep[e.g.,][]{Pont2008,Nikolov2015,Mallonn2016,Rackham2017}.  Natural clouds and hazes might be common in planetary atmospheres and provide a background of ``false positives" against which any detection of artificial climate engineering would be difficult.  One possible distinguishing property of an artificial haze might be an extremely monotonic size distribution $s = s_{\rm haze}$.  While the overall spectral shape, including the break at wavelength $\lambda = 2\pi s_{\it haze}$ of a monotonic distribution of Mie scatterers can be produced by some size distributions, the ``ripple" pattern with wavelength due to interference is unique.    

Not considered here are the questions of {\it whether} or {\it why} an agency would perform such a feat of engineering to maintain the temperature climate on an Earth-like planet.  Self-preservation might be the least of its concerns, since the technical ability needed to construct such long-lived artifacts has probably allowed the agency to leave its planetary cradle behind.  That might not preclude, however, climate stabilization as an act of nostalgia, or of generosity towards future beings that might emerge from that cradle.    

\section*{Acknowledgments}

This manuscript was begun while the author was a visitor at the Center for Space and Habitability at the University of Bern, Switzerland.  The author thanks Ray Pierrehumbert for a stimulating conversation that eventually sparked this work and Daniel Kitzmann for a discussion of hazes.

\end{document}